\documentclass[preprint,showpacs,preprintnumbers,amsmath,amssymb]{revtex4}
\usepackage{graphicx,subfigure}% Include figure files
\usepackage{dcolumn}% Align table columns on decimal point
\usepackage{bm}% bold math
\usepackage{amsmath}
\newcommand{\be}{\begin{equation}}
\newcommand{\ee}{\end{equation}}
\newcommand{\bea}{\begin{eqnarray}}
\newcommand{\eea}{\end{eqnarray}}

\begin{document}

% You should use BibTeX and revtex.bst for references
\bibliographystyle{apsrev}

\preprint{UAB-FT-596}
%\preprint{December 2005}

\title{Unified Model for Inflation and Dark Energy\\
with Planck-Scale Pseudo-Goldstone Bosons} % Force line breaks with \\

\author{Eduard Mass{\'o}}

\author{Gabriel Zsembinszki}

%\thanks{}
\affiliation{Grup de F{\'\i}sica Te{\`o}rica and Institut de
F{\'\i}sica d'Altes
Energies\\Universitat Aut{\`o}noma de Barcelona\\
08193 Bellaterra, Barcelona, Spain}


\date{\today}% It is always \today, today,

\begin{abstract}
We present a model with a complex and a real scalar fields and a
potential whose symmetry is explicitly broken by Planck-scale
physics. For exponentially small breaking, the model accounts for
the period of inflation in the early universe and for the period
of acceleration of the late universe.
\end{abstract}

% , PACS the Physics and Astronomy
\maketitle

\section{Introduction}\label{Introduction}
In spite of the fact that the standard model of elementary
particles based on the gauge group $SU(3)\times SU(2) \times U(1)$
is able to accommodate all existing empirical data, few people
believe that it is the ultimate theory. The reason is that the
standard model leaves unanswered many deep questions. In any case,
evidence (or disproval) of this belief can only be given by
experiment. If we are able to discover a theory that indeed goes
beyond the standard model, it will probably contain new
symmetries. The global symmetries valid at high energies are
expected to be only approximate, since Planck-scale physics
breaks them explicitly \cite{banks,Giddings_etal}. Even if the
effect is probably extremely small, it may lead to very
interesting effects. As has been discussed in \cite{Masso:2004cv},
when a global symmetry is spontaneously broken and we have such a
small explicit breaking the corresponding pseudo-Golstone boson
(PGB) can have a role in cosmology. The focus in
\cite{Masso:2004cv} was to show that the PGB could be a dark
matter constituent candidate.

In the present paper we will rather be concerned with the periods
of acceleration in the universe, namely with inflation in the very
early universe and with dark energy dominance in the late stages
in the evolution of the universe. Recent observational evidence
for these two periods come mainly from the use of Supernovae as
standard candles \cite{supernova}, cosmic microwave background
anisotropies \cite{cmb}, galaxy counts \cite{galaxy} and others
\cite{juan}. Of course, the physics behind the two periods that
are so distant in time may be completely unrelated. However, an
appealing possibility is that they have a common origin. An idea
for this kind of unification has been forwarded by Frieman and
Rosenfeld \cite{Rosenfeld:2005mt}. Their framework is an axion
field model where we have a global $U(1)_{\rm PQ}$ symmetry
spontaneously broken at a high scale and explicitly broken by
instanton effects at the low energy QCD scale. The real part of
the field is able to inflate in the early universe while the axion
boson could be the responsible for the dark energy period. The
authors of that work \cite{Rosenfeld:2005mt} compare their model
of quintessential inflation with other models of inflation and/or
dark energy. We would like to show here that our model of
Planck-scale broken symmetry offers an explicit scenario of a
quintessential inflation.

In our model, we have a complex field $\Psi$ that is charged under
a certain global $U(1)$ symmetry, and in the potential we have the
following $U(1)$-symmetric term
\begin{equation}
V_1(\Psi) =\frac14\lambda[|\Psi|^2-v^2]^2\label{V1}
\end{equation}
where $\lambda$ is a coupling and $v$ is the energy scale of the
spontaneous symmetry breaking (SSB).

We do not need to know the details of how Planck-scale physics breaks
our $U(1)$. It is enough to introduce the most simple effective
 $U(1)$-breaking term
\begin{equation}
 V_{non-sym}(\Psi)=-g\frac1{M_P^{n-3}}|\Psi|^n\left(\Psi
e^{-i\delta}+\Psi^{\star} e^{i\delta}\right)\label{Vnon_sym1}
\end{equation}
($n>3$). Here, $M_P^2\equiv G_N^{-1}$, and the coupling $g$ is
expected to be very small \cite{Kallosh:1995hi}. The term vanishes
when $M_P \rightarrow \infty$, as it should be. Previous work on
explicit breaking of global symmetries can also be found in
\cite{Ross}, and related to Planck-scale breaking, in
\cite{Lusignoli}. Cosmological consequences of some classes of
PGBs are discussed in \cite{Hill}.

Let us write the field as
\begin{equation}
\Psi =\phi\, e^{i\theta/v}\label{Psi}
\end{equation}
We envisage a model where $\phi$, the real part of $\Psi$, is the
inflaton and the PGB $\theta$, the imaginary part of $\Psi$, is a
quintessence field. In the proces of SSB at temperatures $T\sim v$
in the early universe, the scalar field $\phi$ develops in time,
starting from $\phi=0$ and going to values different from zero. A
suitable model we will employ is the inverted hybrid inflation
\cite{Ovrut:1984qp, Lyth:1996kt}, where one has a new real field
$\chi$ that assists $\phi$ to inflate.

The new scalar field is massive and neutral under $U(1)$. We shall
follow ref.\cite{Lyth:1996kt} and couple $\chi$ to $\Psi$ with a
$- \Psi^* \Psi \chi^2$ term. More specifically we introduce the
following contribution to the potential
\begin{equation}
V_2(\Psi,\chi)=\frac12m_{\chi}^2\chi^2+\left(\Lambda^2-
\frac{\alpha^2|\Psi|^2\chi^2}{4\Lambda^2}\right)^2\label{V2}
\end{equation}
Here $\alpha$ is a coupling and $\Lambda$ and $m_{\chi}$ are mass scales.
The interaction between the two fields will give the needed
behavior of the real part of $\Psi$ to give inflation. Also, we should mention
that such models of inflation are realized
in supersymmetry, using a globally supersymmetric scalar potential
\cite{Lyth:1996kt}.

To summarize, our model has a complex field $\Psi$ and a real
field $\chi$ with a total potential
\begin{equation}
V(\Psi,\chi) = V_{sym}(\Psi,\chi) +
V_{non-sym}(\Psi)+C\label{V_tot}
\end{equation}
where $C$ is a constant that sets the minimum of the effective
potential at zero. The non-symmetric part is given by
(\ref{Vnon_sym1}), whereas the symmetric part is the sum of
(\ref{V1}) and (\ref{V2}),
\begin{equation}
V_{sym}(\Psi,\chi) = V_1(\Psi) + V_2(\Psi,\chi)\label{Vsym1}
\end{equation}

\section{The model}
\label{Model}

%In view of the model considered in our previous work (ref PGB), we
%have to apply some modifications on the form of the effective
%potential in order to achieve a reliable model explaining both
%inflation, and the present dark energy of the Universe.
%Specifically, the modified effective potential will contained a
%new auxiliary field, $\chi$, that will couple to the complex
%scalar field $\Psi$ as in hybrid models of inflation
%\cite{Linde}.}
In this section, we will explain in detail the model introduced in
Section \ref{Introduction}. Our basic idea is that the radial part
$\phi$ of the complex scalar field $\Psi$ is responsible for
inflation, whereas the angular part $\theta$ plays the role of the
present dominating dark energy of the universe. From now on, we
will replace $\Psi$ by its expression given in (\ref{Psi}). The
symmetric part of the potential is given by
\begin{equation}
V_{sym}(\phi,\chi)=\frac14\lambda[\phi^2-v^2]^2+
\frac12m_{\chi}^2\chi^2+\left(\Lambda^2-\frac{\alpha^2\phi^2\chi^2}{4\Lambda^2}\right)^2
\label{Vsym}
\end{equation}
while the symmetry-breaking term is
\begin{equation}
V_{non-sym}(\phi,\theta)=
-2\,g\frac{\phi^{n+1}}{M_P^{n-3}}\cos{\frac{\theta}v}
\label{Vnon_sym}
\end{equation}
where the following change of variables
$\theta/v\longrightarrow\theta/v+\delta$ has been made.
%(Couplings of $\chi$ to $\theta$ are not considered due to the
%smallness of the PGB interactions, and couplings of $\chi$ to the
%non-symmetric part of $V$ are neglected?)

\subsection{Inflation}\label{Inflation}
Let us study, firstly, the conditions to be imposed on our model
to describe the inflationary stage of expansion of the primordial
Universe. In order to do this, we will only work with the
symmetric part of the effective potential,
\begin{equation}
V_{sym}(\phi,\chi)=\Lambda^4+\frac12\left(m_{\chi}^2
-\alpha^2\phi^2 \right) \chi^2+
\frac{\alpha^4\phi^4\chi^4}{16\Lambda^4}
+\frac14{\lambda}(\phi^2-v^2)^2, \label{Vsym2}
\end{equation}
which dominates over the non-symmetric one at early times. Here,
$\phi$ is the inflaton field and $\chi$ is the field that plays
the role of an auxiliary field, which is needed to have a sudden
end of the inflationary regime, through a "waterfall" mechanism.
This is important because in this way we can arrange for the right
number of e-folds of inflation and for the right value of the
spectral index of density perturbations produced during inflation,
when the cosmological scales left the horizon. We also note that
the $\phi^4\chi^4$ term in Eq.(\ref{Vsym2}) does not play an
important role during inflation, but only after it, and it sets
the position of the global minimum of $V_{sym}(\phi,\chi)$, which
will roughly be at $\phi\sim v$ and $\chi\sim M$, with
$M=\frac{2\Lambda^2}{\alpha v}$.

The effective mass squared of the field $\chi$ is:
\begin{equation}
M_{\chi}^2=m_{\chi}^2-\alpha^2\phi^2 \label{Mass_of_chi}
\end{equation}
so that for $\phi<\phi_c=\frac{m_{\chi}}{\alpha}$, the only
minimum of $V_{sym}(\phi,\chi)$ is at $\chi=0$. The curvature of
the effective potential in the $\chi$ direction is positive, while
in the $\phi$ direction is negative. This will lead to rollover of
$\phi$, while $\chi$ will stay at its minimum $\chi=0$ until the
curvature in $\chi$ direction changes sign. That happens when
$\phi>\phi_c$ and $\chi$ becomes unstable and starts to roll down
its potential. The mechanism and the conditions to be imposed on
our model are similar to the original hybrid inflation model by
Linde \cite{Linde:1993cn}, so we will follow the same line of
discussion. The main difference between the original model and our
case consists in the fact that here the inflaton rollover is due
to its negative squared mass $m_{\phi}^2=-\lambda v^2$ and it
starts moving from the origin $\phi=0$ towards the minimum
$\langle\phi\rangle\sim v\lesssim M_P$, so that there is no need
to go to values for $\phi$ larger then the Planck scale. The fact
that, initially, the inflaton field is placed at the origin is
justified because in the very hot primordial plasma the symmetry
of the effective potential is restored and the minimum of the
potential is at $\phi=0$. So we expect that, after the SSB, the
radial field $\phi$ is set at the origin of the effective
potential. However, due to quantum fluctuations, the field may be
displaced from $\phi=0$, such that it is unstable and may roll
down the potential.

As is characteristic for hybrid models of inflation
\cite{Lyth:1998xn}, the dominant term in (\ref{Vsym2}) is
$\Lambda^4$. This is equivalent to writing:
\begin{equation}
\frac14\lambda v^4<\Lambda^4. \label{Lambda_Dominant}
\end{equation}
Another requirement is that the absolute mass squared of the
inflaton be much less than the $\chi$-mass squared,
\begin{equation}
|m_{\phi}^2|=\lambda v^2\ll m_{\chi}^2, \label{Small_Infl_Mass}
\end{equation}
which fixes the initial conditions for the fields: $\chi$ is
initially constrained at the stable minimum $\chi=0$, and $\phi$
may slowly roll from its initial position $\phi\simeq 0$.

Taking into account condition (\ref{Lambda_Dominant}), the Hubble
parameter at the time of the phase transition is given by:
\begin{equation}
H^2=\frac{8\pi}{3M_P^2}V_{sym}(\phi_c,0)\simeq\frac{8\pi}{3M_P^2}\Lambda^4.
\label{Hubble_param}
\end{equation}
We want $\phi$ to give sufficient inflation, that is, the
potential $V_{sym}(\phi,0)$ must fulfill the slow-roll conditions
in $\phi$ direction, given by the two slow-roll parameters:
\begin{equation}
\epsilon\equiv\frac{M_P^2}{16\pi}\left(\frac{V'_{sym}}{V_{sym}}\right)^2\ll
1, \label{epsilon}
\end{equation}

\begin{equation}
\arrowvert\eta\arrowvert\equiv\left|\frac{M_P^2}{8\pi}
\frac{V''_{sym}}{V_{sym}}\right|\ll 1 \label{eta}
\end{equation}
where a prime means derivative with respect to $\phi$. The first
slow-roll condition Eq.(\ref{epsilon}) gives
\begin{equation}
\Lambda^4\gg\frac{\lambda}{4\sqrt{\pi}}M_P v^3 \label{slow_roll1}
\end{equation}
and the second slow-roll condition, Eq.(\ref{eta}), gives
\begin{equation}
\Lambda^4\gg\frac{\lambda}{8\pi}M_P^2 v^2. \label{slow_roll2}
\end{equation}
So, under these conditions, the universe undergoes a stage of
inflation at values of $\phi<\phi_c$. In order to calculate the
number of e-folds produced during inflation, we use the following
equation \cite{Lyth:1998xn}
\begin{equation}
N(\phi)=\int_{t}^{t_{end}}H(t)dt=\frac{8\pi}{M_P^2}\int_{\phi_{end}}^{\phi}
\frac{V_{sym}}{V'_{sym}}d\phi\label{N_efolds}
\end{equation}
where $\phi_{end}\equiv\phi(t_{end})=\phi_c$ marks the end of
slow-roll inflation, and prime means derivative with respect to
$\phi$.

Let us study the behavior of the fields $\phi$ and $\chi$ just
after the moment when the field $\phi=\phi_c$ for a period $\Delta
t=H^{-1}=\sqrt{\frac3{8\pi}}\frac{M_P}{\Lambda^2}$. The equation
of motion of the inflaton field, in the slow-roll approximation,
is
\begin{equation}
3H\dot\phi+\frac{\partial V_{sym}(\phi,0)}{\partial \phi}=0.
\label{eq_motion_infl}
\end{equation}
In the time interval $\Delta t=H^{-1}$, the field $\phi$ increases
from $\phi_c$ to $\phi_c+\Delta\phi$. If we suppose that $\phi_c$
takes an intermediate value between 0 and $v$, we can calculate
$\Delta\phi$ using (\ref{eq_motion_infl})
\begin{equation}
3H\frac{\Delta\phi}{\Delta t}\simeq \frac38\lambda v^3
\label{Delta_phi1}
\end{equation}
where, for definiteness, we set $\phi_c\simeq v/2$. We finally get
\begin{equation}
\Delta\phi\simeq\frac3{64\pi}\frac{\lambda v^3M_P^3}{\Lambda^4}.
\label{Delta_phi2}
\end{equation}
The variation of $M_{\chi}^2$ in this time interval is given by
\begin{equation}
\Delta M_{\chi}^2\simeq-\frac3{64\pi}\frac{\lambda \alpha^2 v^4
M_P^2}{\Lambda^4}.\label{Delta_M_chi}
\end{equation}
The field $\chi$ will roll down towards its minimum $\chi_{min}$
much faster than $\phi$, if $|\Delta M_{\chi}^2|\gg H^2$. Taking
into account Eqs.(\ref{Hubble_param}) and (\ref{Delta_M_chi}),
this condition is equivalent to
\begin{equation}
\Lambda^4\ll\frac1{16}\sqrt{\frac{\lambda}{2}}\alpha v^2 M_P^2.
\label{fast_roll_chi}
\end{equation}
In this time interval, $\chi$ rolls down to its minimum,
oscillates around it with decreasing amplitude due to the
expansion of the Universe, and finally stops at the minimum.

Once the auxiliary field $\chi$ arrives and settles down at the
minimum, the inflaton field $\phi$ can roll down towards the
absolute minimum of the potential, much faster than in the case
when $\phi<\phi_c$, because the potential has a non-vanishing
first derivative at that point
\begin{equation}
\frac{\partial V_{sym}}{\partial\phi}
=\lambda\phi(\phi^2-v^2)-\alpha^2 \chi_{min}^2\phi
\label{first_deriv_chi_min}
\end{equation}
which we want to be large in order to assure that no significant
number of e-folds is produced during this part of the field
evolution. The requirement of fast-rolling of $\phi$ is translated
into the following condition
\begin{equation}
v\lesssim M_P \label{fast_roll_phi}
\end{equation}
(this was obtained considering the equation of motion of an
harmonic oscillator with small friction term $3H\dot\phi$, and
imposing the condition that the frequency $\omega^2\gtrsim H^2$).

\subsection{Dark Energy}\label{DE}
Let us now focus on the angular part $\theta$ of the complex
scalar field $\Psi$, which we neglected when discussed about
inflation. We want the PGB $\theta$ to be the field responsible
for the present acceleration of the universe. For this to happen,
we have to impose two conditions on our model: ({\it i}) the field
$\theta$ must be stuck at an arbitrary initial value after the SSB
of $V$, which we suppose is of order $v$, and will only start to
fall towards its minimum in the future; ({\it ii}) the energy
density of the $\theta$ field, $\rho_0$, must be comparable with
the present critical density $\rho_{c_0}$, if we want $\theta$ to
explain all of the dark energy content of our Universe. Conditions
({\it i}) and ({\it ii}) may be written as
\begin{equation}
m_{\theta}\lesssim 3H_0\label{DE_mass_cond}
\end{equation}
\begin{equation}
\rho_{\theta}\sim \rho_{c_0}.\label{DE_energy_cond}
\end{equation}
where $H_0$ is the Hubble constant. Taking into account the
expression for the mass of $\theta$ derived in
\cite{Masso:2004cv},
$m_{\theta}=\sqrt{2g}\left(\frac{v}{M_P}\right)^{\frac{n-1}2}M_P$,
condition (\ref{DE_mass_cond}) becomes
\begin{equation}
g\left(\frac{v}{M_P}\right)^{n-1}\lesssim \frac{9
H_0^2}{2M_P^2}.\label{DE_mass_cond2}
\end{equation}
The energy density of the $\theta$ field is given by the value of
the non-symmetric part of the effective potential,
$V_{non-sym}(\phi,\theta)$, with the assumption that the present
values of both fields are of order $v$
\begin{equation}
\rho_{\theta}\simeq V_{non-sym}(v,v)=
g\left(\frac{v}{M_P}\right)^{n-1} M_P^2 v^2.\label{DE_theta}
\end{equation}
Introducing (\ref{DE_theta}) into (\ref{DE_energy_cond}) and
remembering that the present critical energy density
$\rho_{c_0}=\frac{3H_0^2M_P^2}{8\pi}$, we have that
\begin{equation}
g\left(\frac{v}{M_P}\right)^{n-1}\simeq \frac{3H_0^2}{8\pi
v^2}.\label{DE_energy_cond2}
\end{equation}
Combining (\ref{DE_mass_cond2}) and (\ref{DE_energy_cond2}) we get
\begin{equation}
\frac{3H_0^2}{8\pi v^2}\lesssim
\frac{9H_0^2}{2M_P^2}\label{DE_cond1}
\end{equation}
which finally gives
\begin{equation}
v\gtrsim \frac16 M_P.\label{DE_cond2}
\end{equation}
This is the restriction to be imposed on $v$ in order for $\theta$
to be the field describing dark energy. Notice that it is
independent of $n$. It is also interesting to obtain the
restriction on the coupling $g$, which can be done if we introduce
(\ref{DE_cond2}) into (\ref{DE_energy_cond2}) giving
\begin{equation}
g\lesssim \frac{3\times
6^{n+1}}{8\pi}\frac{H_0^2}{M_P^2}.\label{g_cond1}
\end{equation}
Replacing the value for $H_0\sim 10^{-42}$ GeV and taking the
smallest value $n=4$, we obtain the limit
\begin{equation}
g\lesssim 10^{-119}.\label{g_cond2}
\end{equation}

\section{Discussions and Conclusions}\label{Conclusions}
\begin{figure}[htb]
\begin{center}
\includegraphics[width=9cm, height=7cm]{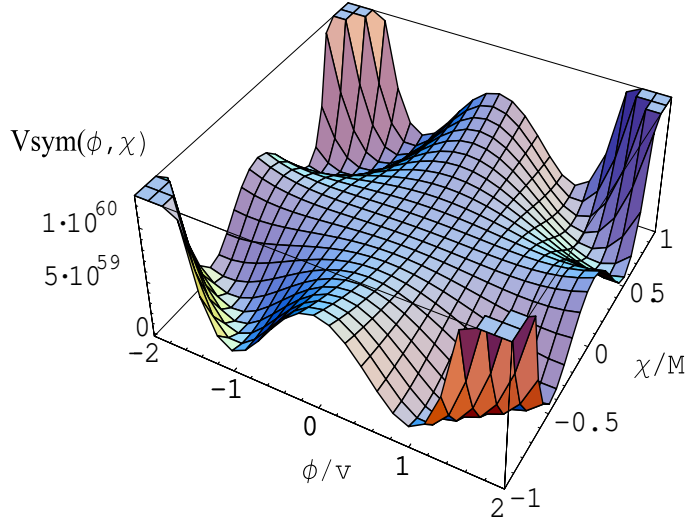}
\end{center}
\caption{The symmetric part of the effective potential, $V_{sym}$,
as function of the two normalized fields $(\phi/v,
\chi/M)$}\label{fig_potential}
\end{figure}

\begin{figure}[htb]
\begin{center}
\includegraphics[width=10cm, height=5.5cm]{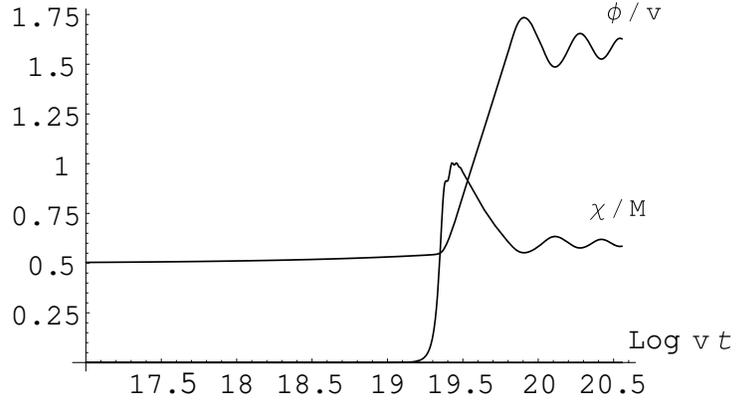}
\end{center}
\caption{Numerical solution for the system of the coupled
equations of motion of the two fields, $\tilde\phi=\phi/v$ and
$\tilde\chi=\chi/M$. Notice the logarithmic time-scale on the
abscise}\label{fields_evol}
\end{figure}

In the previous section, we derived the conditions to be imposed
on the parameters of our model in order to give the right
description for inflation and dark energy. Let us give here a
numerical example and show the field evolution. In all the
figures, we use the following values for the parameters:
$v=0.5\times 10^{19}$ GeV, $\lambda=10^{-16}$, $\Lambda=9\times
10^{14}$ GeV, $m_{\chi}=2.5\times 10^{12}$ and $\alpha=10^{-6}$.
The tiny value for $\lambda$ is needed in order to generate the
correct amplitude of density perturbations, $\delta \rho/\rho\sim
2\times 10^{-5}$ \cite{KolbTurner}. In Fig.\ref{fig_potential} we
display the graphical representation of the symmetric part
$V_{sym}(\phi,\chi)$ of the effective potential and in
Fig.\ref{fields_evol} we show the numerical solution to the system
of the two equations of motion of the fields $\tilde\phi=\phi/v$
and $\tilde\chi=\chi/M$. We have solved it for the interesting
region starting from $\phi=\phi_c$ and $\chi=0$ (because during
inflation we know that $\chi=0$ and $\phi$ slowly rolls down the
potential). We notice that when $\chi$ approaches its minimum, the
slow-rolling of $\phi$ ceases, and it rapidly evolves towards the
minimum of the effective potential and oscillates around it.

\begin{figure}[htb]
\begin{center}
\includegraphics[width=9.5cm, height=6.5cm]{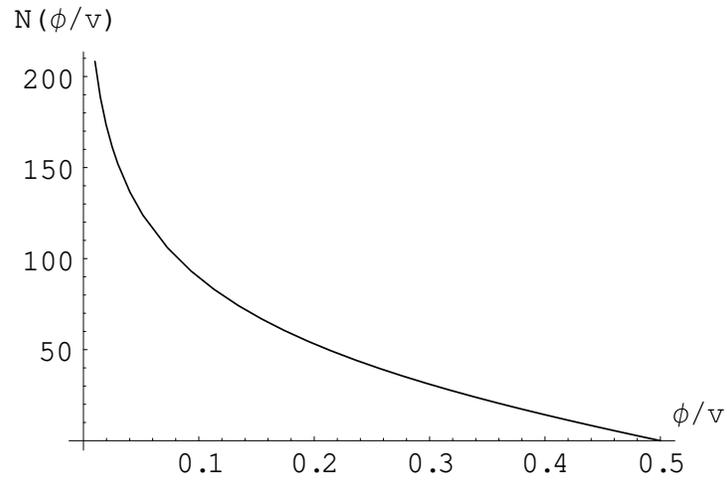}
\end{center}
\caption{The number $N(\tilde\phi)$ of e-folds of inflation as a
function of $\tilde\phi=\phi/v$}\label{Fig_Ne}
\end{figure}

In Fig.\ref{Fig_Ne}, we plot the number of e-folds $N(\phi)$
defined in Eq.(\ref{N_efolds}), as a function of the inflaton
field. The maximum value for $\phi$ that we chose is $\phi_c$,
while the minimum value is $\simeq 0$. The interesting region is
the one that gives "observable inflation", that is for values of
$\phi$ that give $N(\phi)\lesssim 60$. This is because
$N(\phi)\sim 60$ corresponds to the time when cosmological scales
leave the horizon during inflation. All what happened before is
outside our horizon and is totally irrelevant at the present time.

\begin{figure}[]
  \centering
  \subfigure[]{
    \includegraphics[width=9.5cm, height=5.5cm]{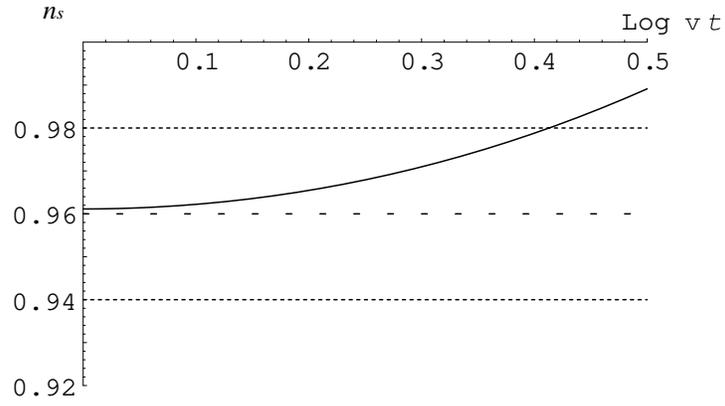}
    \label{fig1}}

  \subfigure[]{
    \includegraphics[width=10cm, height=5cm]{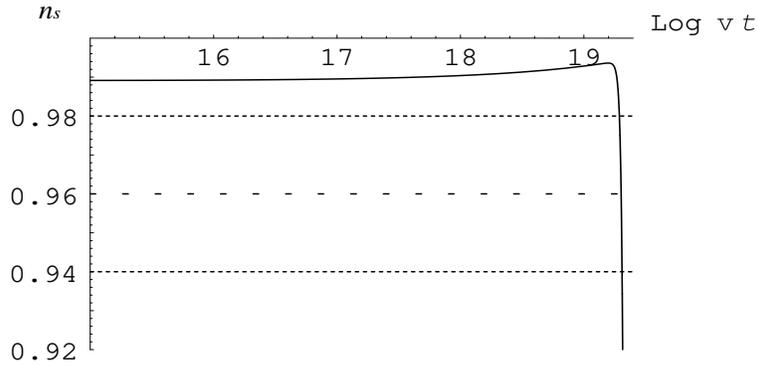}
    \label{fig2}}
\caption{$(a)$ The spectral index $n_s$ during inflation, as a
function of the inflaton field, and $(b)$ just after the time when
$\phi=\phi_c$, as a function of the logarithm of time. In the
above figures, the dashed line is the expected value for $n_s$
within experimental errors, delimited by the pointed lines.}
\label{Fig_ns}
\end{figure}

A way to confront the predictions of our model with observational
data is through the spectral index $n_s$ of density perturbations
produced during inflation \cite{juan,Alabidi:2005qi}. The spectral
index is defined in terms of the slow-roll parameters $\epsilon$
and $\eta$ by the relation
\begin{equation}
n_s=1+2\eta-6\epsilon\label{spectral_index}
\end{equation}
and experimental data indicate a value of $n_s=0.96\pm 0.02$
\cite{cmb}. We display in Fig.\ref{Fig_ns} the dependence of the
spectral index on the inflaton field $\phi$.

One of our numerical conclusions is that $g$ has to have an
extremely small value, as we see in (\ref{g_cond2}). It says that
the effect of Planck-scale physics in breaking global symmetries
should be exponentially suppressed. Let us mention at this point
that there are arguments for such a strong suppression. Indeed,
interest in the quantification of the effect came from the fact
that the consequence of the explicit breaking of the Peccei-Quinn
symmetry \cite{Peccei:1977hh} is that the Peccei-Quinn mechanism
is no longer a solution to the strong CP-problem
\cite{Holman:1992us}.

In ref. \cite{Kallosh:1995hi} it was shown that in string-inspired
models there could be non-perturbative symmetry breaking effects
of order $\exp{\left[- \pi (M_P/M_{\rm string})^2\right]}$. For
$M_{\rm string}<10^{18}$ GeV, we get (\ref{g_cond2}). Although we
considered perturbative effects, that analysis shows that perhaps
the values (\ref{g_cond2}) leading to $\theta$ being quintessence
are realistic.

Finally we would like to comment on the possibility that instead
of having one field $\Psi$ we have $N$ fields $\Psi_1$, $\Psi_2$,
... $\Psi_N$. We are motivated by the recent work
\cite{Dimopoulos:2005ac} where $N$ inflatons are introduced. The
interesting case is when $N$ is large, as suggested in some
scenarios discussed in \cite{Dimopoulos:2005ac}. When having $N$
fields, our relations should of course be modified. In the simple
case that the parameters of the $N$ fields are identical, to
convert the formulae in the text to the new case, we should make
the following changes: $v \rightarrow N^{1/2} v$, $\Lambda
\rightarrow N^{1/4}\Lambda$, $g\rightarrow N^{-3/2} g$ and
$\lambda\rightarrow N^{-1}\lambda$. This would allow to change the
values of the parameters, for example with large values for $N$ we
can have smaller values for $v$, and so on.

To summarize, our purpose has been to give a step forward starting
from the idea of Frieman and Rosenfeld \cite{Rosenfeld:2005mt}
that fields in a potential may supply a unified explanation of
inflation and dark energy. Our model contains two scalar fields,
one complex and one real, and a potential that contains a
non-symmetric part due to Planck-scale physics. We have determined
the conditions under which our fields can act as inflaton and as
quintessence. One of the conditions is that the explicit breaking
has to be exponentially suppressed, as suggested by quantitative
studies of the breaking of global symmetries by gravitational
effects \cite{Kallosh:1995hi}.

\begin{acknowledgments}
We are grateful to F. Rota for his participation in the first
stages of this work. We acknowledge support by the CICYT Research
Project FPA2005-05904 and by the \textit{Departament
d'Universitats, Recerca i Societat de la Informaci{\'o}} (DURSI),
Project 2005SGR00916. The work of G.Z. is supported by the DURSI,
under grant 2003FI 00138.
\end{acknowledgments}


\begin{thebibliography}{99}

%\cite{banks}
\bibitem{banks}
See for instance T.~Banks, Physicalia {\bf 12}, 19 (1990), and
references therein.


%\cite{Giddings_etal}
\bibitem{Giddings_etal}
S.~B.~Giddings and A.~Strominger,
 %``Loss Of Incoherence And Determination Of Coupling Constants In Quantum
%Gravity,''
Nucl.\ Phys.\ B {\bf 307}, 854 (1988).
%%CITATION = NUPHA,B307,854;%%
\\
S.~R.~Coleman,
 %``Why There Is Nothing Rather Than Something: A Theory Of The Cosmological
%Constant,''
Nucl.\ Phys.\ B {\bf 310}, 643 (1988).
%%CITATION = NUPHA,B310,643;%%
\\
G.~Gilbert,
%``Wormhole Induced Proton Decay,''
Nucl.\ Phys.\ B {\bf 328}, 159 (1989).
%%CITATION = NUPHA,B328,159;%%

%\cite{Masso:2004cv}
\bibitem{Masso:2004cv}
  E.~Masso, F.~Rota and G.~Zsembinszki,
  %``Planck-scale effects on global symmetries: Cosmology of  pseudo-Goldstone
  %bosons,''
  Phys.\ Rev.\ D {\bf 70}, 115009 (2004)
  [arXiv:hep-ph/0404289].
  %%CITATION = HEP-PH 0404289;%%

\bibitem{supernova}
S.~Perlmutter {\it et al.}  [Supernova Cosmology Project
Collaboration],
%``Measurements of Omega and Lambda from 42 High-Redshift Supernovae,''
Astrophys.\ J.\  {\bf 517}, 565 (1999) [arXiv:astro-ph/9812133].
\\
A.~G.~Riess {\it et al.}  [Supernova Search Team Collaboration],
%``Observational Evidence from Supernovae for an Accelerating Universe and a
%Cosmological Constant,''
Astron.\ J.\  {\bf 116}, 1009 (1998) [arXiv:astro-ph/9805201].

\bibitem{cmb}
  D.~N.~Spergel {\it et al.}  [WMAP Collaboration],
  %``First Year Wilkinson Microwave Anisotropy Probe (WMAP) Observations:
  %Determination of Cosmological Parameters,''
  Astrophys.\ J.\ Suppl.\  {\bf 148}, 175 (2003)
  [arXiv:astro-ph/0302209].
  %%CITATION = ASTRO-PH 0302209;%%
\\
A.~Balbi {\it et al.},
%``Constraints on cosmological parameters from MAXIMA-1,''
Astrophys.\ J.\  {\bf 545}, L1 (2000) [Erratum-ibid.\  {\bf 558},
L145 (2001)] [arXiv:astro-ph/0005124].
\\
C.~Pryke, N.~W.~Halverson, E.~M.~Leitch, J.~Kovac,
J.~E.~Carlstrom, W.~L.~Holzapfel and M.~Dragovan,
%``Cosmological Parameter Extraction from the First Season of Observations with
%DASI,''
Astrophys.\ J.\  {\bf 568}, 46 (2002) [arXiv:astro-ph/0104490].
\\
C.~B.~Netterfield {\it et al.}  [Boomerang Collaboration],
%``A measurement by BOOMERANG of multiple peaks in the angular power spectrum
%of the cosmic microwave background,''
Astrophys.\ J.\  {\bf 571} (2002) 604 [arXiv:astro-ph/0104460].
\\
J.~L.~Sievers {\it et al.},
%``Cosmological Parameters from Cosmic Background Imager Observations and
%Comparisons with BOOMERANG, DASI, and MAXIMA,''
Astrophys.\ J.\  {\bf 591}, 599 (2003) [arXiv:astro-ph/0205387].
\\
A.~Benoit {\it et al.}  [the Archeops Collaboration],
%``Cosmological constraints from Archeops,''
Astron.\ Astrophys.\  {\bf 399}, L25 (2003)
[arXiv:astro-ph/0210306].
\\
%\cite{MacTavish:2005yk}
%\bibitem{MacTavish:2005yk}
  C.~J.~MacTavish {\it et al.},
  %``Cosmological parameters from the 2003 flight of BOOMERANG,''
  arXiv:astro-ph/0507503.
  %%CITATION = ASTRO-PH 0507503;%%
\\
%\cite{Sanchez:2005pi}
%\bibitem{Sanchez:2005pi}
  A.~G.~Sanchez {\it et al.},
  %``Cosmological parameters from CMB measurements and the final 2dFGRS power
  %spectrum,''
  Mon.\ Not.\ Roy.\ Astron.\ Soc.\  {\bf 366}, 189 (2006)
  [arXiv:astro-ph/0507583].
  %%CITATION = ASTRO-PH 0507583;%%


\bibitem{galaxy}
L.~Verde {\it et al.}, Mon.\ Not.\ Roy.\ Astron.\ Soc.\ 335 (2002)
432 [arXiv:astro-ph/0112161].
\\
M.~S.~Turner,
%The Case for \Omega_M = 0.33 ± 0.035
The Astrophysical Journal, 576:L101-L104, 2002.
\\
A.~Lewis and S.~Bridle,
%``Cosmological parameters from CMB and other data: a Monte-Carlo approach,''
Phys.\ Rev.\ D {\bf 66}, 103511 (2002) [arXiv:astro-ph/0205436].
\\
X.~m.~Wang, M.~Tegmark and M.~Zaldarriaga,
%``Is cosmology consistent?,''
Phys.\ Rev.\ D {\bf 65}, 123001 (2002) [arXiv:astro-ph/0105091].

\bibitem{juan}
  J.~Garcia-Bellido,
  %``Cosmology and Astrophysics,''
  arXiv:astro-ph/0502139.


%\cite{Rosenfeld:2005mt}
\bibitem{Rosenfeld:2005mt}
  R.~Rosenfeld and J.~A.~Frieman,
  %``A simple model for quintessential inflation,''
  JCAP {\bf 0509}, 003 (2005)
  [arXiv:astro-ph/0504191].
  %%CITATION = ASTRO-PH 0504191;%%

%\cite{Kallosh:1995hi}
\bibitem{Kallosh:1995hi}
  R.~Kallosh, A.~D.~Linde, D.~A.~Linde and L.~Susskind,
  %``Gravity and global symmetries,''
  Phys.\ Rev.\ D {\bf 52}, 912 (1995)
  [arXiv:hep-th/9502069].
  %%CITATION = HEP-TH 9502069;%%

\bibitem{Ross}
C.~T.~Hill and G.~G.~Ross,
%``Pseudogoldstone Bosons And New Macroscopic Forces,''
Phys.\ Lett.\ B {\bf 203}, 125 (1988).
%%CITATION = PHLTA,B203,125;%%
\\
C.~T.~Hill and G.~G.~Ross,
 %``Models And New Phenomenological Implications Of A Class Of Pseudogoldstone
%Bosons,''
Nucl.\ Phys.\ B {\bf 311}, 253 (1988).
%%CITATION = NUPHA,B311,253;%%


\bibitem{Lusignoli}
M.~Lusignoli, A.~Masiero and M.~Roncadelli,
%``Spontaneous Versus Explicit Breaking Of A Continuous Global Symmetry,''
Phys.\ Lett.\ B {\bf 252}, 247 (1990).
%%CITATION = PHLTA,B252,247;%%
\\
S.~Ghigna, M.~Lusignoli and M.~Roncadelli,
%``Instability of the invisible axion,''
Phys.\ Lett.\ B {\bf 283}, 278 (1992).
%%CITATION = PHLTA,B283,278;%%
\\
D.~Grasso, M.~Lusignoli and M.~Roncadelli,
%``Global continuous symmetry and the 17-keV neutrino,''
Phys.\ Lett.\ B {\bf 288}, 140 (1992).
%%CITATION = PHLTA,B288,140;%%


\bibitem{Hill}
C.~T.~Hill, D.~N.~Schramm and J.~N.~Fry,
%``Cosmological Structure Formation From Soft Topological Defects,''
Comments Nucl.\ Part.\ Phys.\  {\bf 19}, 25 (1989).
%%CITATION = CNPPA,19,25;%%
\\
A.~K.~Gupta, C.~T.~Hill, R.~Holman and E.~W.~Kolb,
%``Statistical mechanics of soft boson phase transitions,''
Phys.\ Rev.\ D {\bf 45}, 441 (1992).
%%CITATION = PHRVA,D45,441;%%
\\
J.~A.~Frieman, C.~T.~Hill and R.~Watkins,
 %``Late time cosmological phase transitions. 1. Particle physics models and
%cosmic evolution,''
Phys.\ Rev.\ D {\bf 46}, 1226 (1992).
%%CITATION = PHRVA,D46,1226;%%
\\
J.~A.~Frieman, C.~T.~Hill, A.~Stebbins and I.~Waga,
%``Cosmology with ultralight pseudo Nambu-Goldstone bosons,''
Phys.\ Rev.\ Lett.\  {\bf 75}, 2077 (1995)
[arXiv:astro-ph/9505060].
%%CITATION = ASTRO-PH 9505060;%%


%\cite{Ovrut:1984qp}
\bibitem{Ovrut:1984qp}
  B.~A.~Ovrut and P.~J.~Steinhardt,
  %``Inflationary Cosmology And The Mass Hierarchy In Locally Supersymmetric
  %Theories,''
  Phys.\ Rev.\ Lett.\  {\bf 53}, 732 (1984).
  %%CITATION = PRLTA,53,732;%%


%\cite{Lyth:1996kt}
\bibitem{Lyth:1996kt}
  D.~H.~Lyth and E.~D.~Stewart,
  %``More varieties of hybrid inflation,''
  Phys.\ Rev.\ D {\bf 54}, 7186 (1996)
  [arXiv:hep-ph/9606412].
  %%CITATION = HEP-PH 9606412;%%


%\cite{Linde:1993cn}
\bibitem{Linde:1993cn}
  A.~D.~Linde,
  %``Hybrid inflation,''
  Phys.\ Rev.\ D {\bf 49}, 748 (1994)
  [arXiv:astro-ph/9307002].
  %%CITATION = ASTRO-PH 9307002;%%

%\cite{Lyth:1998xn}
\bibitem{Lyth:1998xn}
  D.~H.~Lyth and A.~Riotto,
  %``Particle physics models of inflation and the cosmological density
  %perturbation,''
  Phys.\ Rept.\  {\bf 314}, 1 (1999)
  [arXiv:hep-ph/9807278].
  %%CITATION = HEP-PH 9807278;%%
\\
  E.~J.~Copeland, A.~R.~Liddle, D.~H.~Lyth, E.~D.~Stewart and D.~Wands,
  %``False vacuum inflation with Einstein gravity,''
  Phys.\ Rev.\ D {\bf 49}, 6410 (1994)
  [arXiv:astro-ph/9401011].
  %%CITATION = ASTRO-PH 9401011;%%




\bibitem{KolbTurner}
E.~W.~Kolb and M.~S.~Turner, ``The Early Universe,'' {\it  Redwood
City, USA: Addison-Wesley (1990) 547 p. (Frontiers in physics,
69)}.
\\
  E.~F.~Bunn, A.~R.~Liddle and M.~J.~White,
  %``Four-year COBE normalization of inflationary cosmologies,''
  Phys.\ Rev.\ D {\bf 54}, 5917 (1996)
  [arXiv:astro-ph/9607038].
  %%CITATION = ASTRO-PH 9607038;%%

%\cite{Alabidi:2005qi}
\bibitem{Alabidi:2005qi}
  L.~Alabidi and D.~H.~Lyth,
  %``Inflation models and observation,''
  arXiv:astro-ph/0510441.
  %%CITATION = ASTRO-PH 0510441;%%

%\cite{Peccei:1977hh}
\bibitem{Peccei:1977hh}
  R.~D.~Peccei and H.~R.~Quinn,
  %``CP Conservation In The Presence Of Instantons,''
  Phys.\ Rev.\ Lett.\  {\bf 38} (1977) 1440.
  %%CITATION = PRLTA,38,1440;%%
\\
R.~D.~Peccei and H.~R.~Quinn,
  %``Constraints Imposed By CP Conservation In The Presence Of Instantons,''
  Phys.\ Rev.\ D {\bf 16} (1977) 1791.
  %%CITATION = PHRVA,D16,1791;%%

%\cite{Holman:1992us}
\bibitem{Holman:1992us}
  R.~Holman, S.~D.~H.~Hsu, T.~W.~Kephart, E.~W.~Kolb, R.~Watkins and L.~M.~Widrow,
  %``Solutions to the strong CP problem in a world with gravity,''
  Phys.\ Lett.\ B {\bf 282} (1992) 132
  [arXiv:hep-ph/9203206].
  %%CITATION = HEP-PH 9203206;%%
\\
%\cite{Kamionkowski:1992mf}
%\bibitem{Kamionkowski:1992mf}
M.~Kamionkowski and J.~March-Russell,
  %``Planck scale physics and the Peccei-Quinn mechanism,''
  Phys.\ Lett.\ B {\bf 282} (1992) 137
  [arXiv:hep-th/9202003].
  %%CITATION = HEP-TH 9202003;%%
\\
%\cite{Barr:1992qq}
%\bibitem{Barr:1992qq}
  S.~M.~Barr and D.~Seckel,
  %``Planck scale corrections to axion models,''
  Phys.\ Rev.\ D {\bf 46} (1992) 539.
  %%CITATION = PHRVA,D46,539;%%


%\cite{Dimopoulos:2005ac}
\bibitem{Dimopoulos:2005ac}
  S.~Dimopoulos, S.~Kachru, J.~McGreevy and J.~G.~Wacker,
  %``N-flation,''
  arXiv:hep-th/0507205.
  %%CITATION = HEP-TH 0507205;%%



\end{thebibliography}

\newpage

\end{document}